# Linear Programming Approaches for Power Savings in Software-defined Networks (The Extended Version)

Fahimeh Alizadeh Moghaddam
SNE & S2 groups
University of Amsterdam & VU University Amsterdam
Amsterdam, The Netherlands
Email: f.alizadehmoghaddam@uva.nl

Paola Grosso
System and Network Engineering group (SNE)
University of Amsterdam
Amsterdam, The Netherlands
Email: p.grosso@uva.nl

*Abstract*—Software-defined networks have been proposed as a viable solution to decrease the power consumption of the networking component in data center networks. Still the question remains on which scheduling algorithms are most suited to achieve this goal. We propose 4 different linear programming approaches that schedule requested traffic flows on SDN switches according to different objectives. Depending on pre-defined software quality requirements such as delay and performance, a single variation or a combination of variations can be selected to optimize the power saving and the performance metrics. Our simulation results demonstrate that all our algorithm variations outperform the shortest path scheduling algorithm, our baseline on power savings, less or more strongly depending on the power model chosen. We show that in FatTree networks, where switches can save up to 60% of power in sleeping mode, we can achieve 15% minimum improvement assuming a one-to-one traffic scenario. Two of our algorithm variations privilege performance over power saving and still provide around 45% of the maximum achievable savings.

## I. INTRODUCTION

Networking devices are one of the main contributors to power consumption in data centers. However, improving on their overall power efficiency is not a trivial task. In fact, from the data center providers perspective power efficiency in the network needs to be weighted against other quality requirements such as its performance, reliability and availability. This results in the deployment of large number of switches, connected via multiple (including back-up) links, whose average utilization rate is around 30% [1]. In this context there is room for improvement in the overall energy efficiency of the networking component.

Nowadays, software-defined networks (SDN) are replacing traditional networks due to their advantage of providing programmability and controllability of the underlying network. As Infonetics Research forecasts [2], there will be 87% growth in SDN deployment in North American-based enterprises by 2016. Given SDNs wide range of applications and their rapid growth in data center networks, they are often chosen as part of the power efficient solutions. In an earlier systematic literature review [3], we studied the existing energy efficient networking solutions and we observed a growing trend in using SDNs combined with energy awareness. The existing energy efficient networking solutions are implemented as flow schedulers that fullfill user-defined quality requirements while considering energy consumption improvements. However, the effectiveness of SDNs with respect to energy efficiency are not fully studied and it is not determined how the programmability of networks can be a added value in this matter. Besides, scalability and performance cost of energy efficiency optimization are still open questions.

Linear programming algorithms are recently deployed to perform as a scheduler for the incoming flows on the available paths [4]–[10]. Finding the optimum solution, e.g. mapping the flows to the paths, is an NP-hard problem. In this paper, we will show it is possible to reduce the complexity by splitting and adjusting the problem. We propose 4 linear programming algorithms that differ in their objective functions. We study each algorithm to see to what extent they impact other quality requirements such as performance. Our main contributions are:

- We derive four different flow scheduling algorithms that take into account a number of constraints to provide either highest throughput or highest power efficiency.
- We implement a modular decision framework to evaluate our algorithm variations, which collects statistical information from the network and schedules the existing/new flows accordingly.
- We support our hypothesis with a number of simulation experiments and we compare our results with a baseline, namely the shortest-path scheduling algorithm.
- We assess to what extent the scalability quality requirement is fulfilled by performing experiments in network of different sizes, different numbers of flows, and different characteristics of switches.

The rest of the article is structured as follows. Existing related work on power efficiency solutions for software-defined networks is discussed in section II. In section III, we present our scheduling algorithm and its implementation. We describe our evaluation decision framework and its components in section IV. Sections V and VI present the simulation scenarios and experimental results, respectively. Our findings and observations are discussed in section VII. Finally, the paper is concluded and directions for future work are introduced in section VIII.

## II. RELATED WORK

Power and energy efficiency of data center networks has often been a by-product of other optimization strategies. The main focus in data centers has been on energy-aware virtual machine (VM) placement, and the underlying network is treated as the side problem in terms of providing power savings. In this case, power savings in the network usually are achieved by traffic consolidation or traffic locality [4], [11]. For example, the frameworks introduced in [12], [13], minimize the number of active racks in the data center networks by consolidating VMs in fewer number of racks. VM placement has been identified as a routing problem in [6] and it has been combined with the network optimization. Solely focused on the networking component, [14] introduced a distributed flow scheduling scheme suitable within a data center network. Still, there are no complete studies made on the trade-off between the optimization in the network component and the application related objectives. We instead focus specifically on this, with the goal of providing application developers that intend to program the network a clear overview of the pros and cons of their algorithms choices.

To provide this information to application developers we focus on the FatTree topology, a very well known three-tiered data center network architecture. This isn't the only possible choice: Fang, *et al.* in [15] study different network architectures other than FatTree, namely, 2N-Tree, VL2 and BCube from the power saving point of view. L. Gyarmati, *et al.* design a comparison study on energy consumption of BCube, DCell, FatTree and balancedTree network architectures [16]. As we will discuss in Sec. IV-A the decision framework we developed allows to define arbitrary topologies as input, and as such will allow comparisons between different data center architectures.

There are other studies that, like ours, deploy software-defined networks in order to apply the energy-aware changes decided by their decision frameworks or applications. The pioneer work in this era, ElasticTree [7] designs a centralized decision framework to turn off the idle network devices. A disadvantage of this model is that the multi-minute booting time of switches makes handling appropriately bursty traffic more difficult. Differently, we use in our simulation the sleeping mode of switches; this takes much shorter time (around 1s) to turn them back on, while still providing significant amount of power saving. There are similar approaches that put idle network devices into the sleeping mode or turn them off by implementing heuristic scheduling algorithms [5], [17], [18]. They mostly provide the traffic matrix as an input to the scheduling algorithm, which is not always a realistic scenario. Contrarily, we focus on real time flow requests while keeping the scalability quality requirement in mind, as we will discuss in Sec. III-D.

## III. THE FLOW SCHEDULING ALGORITHM

To enable communication between nodes in an SDN, a scheduling algorithm need to select the route first. Such an algorithm can be energy-agnostic and update the flow tables of the switches, regardless of energy consumption of the route and its throughput. Shortest-path finding approach (SP) is such an example as it does not retrieve any bandwidth/energy related information from the system and consequently can not make smart decisions based this. We use it and its extension as a baseline to compare the quality metrics of our algorithm with. In order to design our power-savvy scheduling algorithm, we deploy the Integer Linear Programming model, which defines the problem as a linear function of different variables and the values selected for variables need to meet the limitations of pre-defined constraints. We aim to schedule the incoming network traffic with power saving intentions and reducing the volume of the traffic by intelligent endpoint placements is out of scope of this research. Therefore, we assume the endpoints are known beforehand. Also, no initial traffic matrix is assigned to the data center and the incoming load is routed in realtime. This makes scalability a very important quality requirement in our scheduling algorithm design.

### A. Objectives

There are three objectives that can be considered when scheduling traffic flows:

- **Minimize the power consumption of the data center network:**
  To minimize the total power consumption, there are multiple options. One is to turn off the idle network devices, but this takes a significant amount of time for the devices to boot and be back up again. Another option is to turn off the idle switch ports separately, when needed. This is in essence the approach taken by techniques like Energy Efficient Ethernet, which are shown to provide some energy savings. Still the device chassis is the main energy consuming component; [7] states that switches consume up to 90% of their maximum power as soon as they are turned on, before any incoming load. Therefore, we place our focus on putting the idle network devices into the sleeping mode. Switches in the sleeping mode still consume energy but much less than when they do in the active mode. This value will depend on the technologies adopted. Besides, their transition time is significantly less than the booting duration. Ultimately, we aim to minimize the number of active switches for a given number of flows in the network.
  Eq. 1 defines the algorithm objective, with total power consumption $PC$ being the sum of the power consumption of each active switch, $m$ the number of active switches selected to serve $n$ number of flows.

$$Minimize(PC = \sum_{j=1}^{m} PC_j) \quad (1)$$

- **Minimize the number of transitions from the sleeping mode to the active mode:**
  Another energy-aware approach that we apply is prioritizing the already existing active switches over the sleeping ones. This help us save the transition time. Consequently, the duration of the scheduling phase will be shorter although the achieved bandwidths for the flows might be reduced due to link sharing. Therefore, it is up to our optimization algorithm to keep the balance between the bandwidths and the power consumption by turning on the switches. We associate this concept with a new variable called *transition_degree*, which indicates the number of switches that need to be turned on for the incoming load.

The *transition_degree* can have a value between 0 and the total number of switches. The objective expressed in Eq. 2 is to minimize this variable.

$$Minimize(transition\_degree) \quad (2)$$

- **Maximize the bandwidth for the flows**
  From the user perspective, the main quality requirement is application performance. In real life scenarios, different applications might have different performance requirements and they will get different priorities. In this work, we assume that all the applications, hence the flows we schedule, have the same priority in regard to throughput and we try to optimize the bandwidth for the flows with fairness.

$$Maximize(\sum_{i=1}^{n} BW_i) \quad (3)$$

Eq. 3 defines the objective, with *n* being the number of flows in the data center network.

### B. Algorithm Variations

Variations in application-level quality requirements might privilege one or more of the objectives. This leads us define different variations of our scheduling algorithm accordingly. For each variation, different objectives are bound together to formulate an objective variable. However, it is important to note that all the variations are power efficient as the idle switches are put into the sleeping mode.

We identified four distinct objective variables:

LP-v1: *Full version*
As Eq. 4 shows, all the objectives are taken into account, where *n* is the number of flows and *m* is the number of active switches. In order to avoid zero denominator in the fraction, 1 is added to *transition_degree*.

$$\frac{\sum_{i=1}^{n} BW_i}{(transition\_degree + 1) * \sum_{j=1}^{m} PC_j} \quad (4)$$

LP-v2: *Without priority*
In this version, all the switches either in the active mode or in the sleeping mode have the same chance of being selected for the next coming flow. We do not give priority to the already active switches as Eq. 5 displays:

$$\frac{\sum_{i=1}^{n} BW_i}{\sum_{j=1}^{m} PC_j} \quad (5)$$

LP-v3: *Only throughput guaranteed*
For delay-sensitive applications, degrading throughput for the purpose of energy efficiency is not acceptable. Therefore, only performance improvements are taken into account into this objective variables, which does not shape the traffic to increase the number of idle devices. This version still achieves some power efficiency by putting the idle switches into sleeping mode if there is any. Eq. 6 shows that in this version we only consider bandwidth values.

$$\sum_{i=1}^{n} BW_i \quad (6)$$

LP-v4: *Only energy efficient*
Our last variation only takes the power efficiency metric into account and minimizes the total power consumption of the data center network, when serving a specific number of flows. To be more efficient, we also include *transition_degree*, which is shown in Eq. 7.

$$(transition\_degree + 1) * \sum_{j=1}^{m} PC_j \quad (7)$$

### C. Algorithm Implementation

In order to model our linear programming problem we define *combination*. A *combination* consists of a set of selected paths for the requested flows. We keep a list of *combinations*, which are distinct and differ in the selected paths and consequently in the achieved bandwidths and the total power consumption. As new flows request arrive, the combinations will grow both in size and in number. Unlike the growth in size, which is adding only one flow and one selected path, the growth in number is more accelerated. Eq. 8 calculates the total number of combinations, where *n* is the number of requested flows. For scalability reasons that will be discussed later in section III-D, we reduce the number of combinations to only three.

$$\text{Number of combinations} = \prod_{i=1}^{n} \text{number of possible paths}_i \quad (8)$$

Each *combination* could be a candidate for providing the paths for requested flows. The pseudo-code in Algorithm 1 describes the steps performed in our scheduling process for each of the algorithm variations. *Combination* selection is done by comparing them based on their objective variable. We calculate the necessary metrics (the maximum power consumption, the maximum bandwidths or *transition_degree*) for each combination. For *l* number of combinations in Eq. 9, we model our linear programming objective function with a list of boolean variables (*x*), which shows if a *combination* is selected, multiplied by the objective variables. In case of variations LP-v1, LP-v2 and LP-v3, the objective function is to maximize the objective variables, whereas it minimizes the objective variable in case of variation LP-v4.

$$\sum_{k=1}^{l} x_k * (\text{Combination objective variable}_k) \quad (9)$$

We define the constraints ($\sum_{k=1}^{l} x_k = 3$ and $\sum_{k=1}^{l} x_k = 1$) to ensure that top 3 combinations are selected for *CombinationList* and only 1 combination is selected as *SelectedCombination*.

**Algorithm 1** Scalable Linear Programming Flow Scheduler
```
while TRUE do
    if newRequestedFlow not Null then
        possiblePaths ← List of possible paths in the data center network
        if CombinationList not ∅ then
            CombinationList ← Update each combination with new possible paths
        else
            CombinationList ← Create a combination with each new possible path
        end if
        for all combination ∈ CombinationList do
            Update the objective variable:
                ←if necessary estimate PC
                ←if necessary estimate BW
                ←if necessary calculate transition_degree
        end for
        CombinationList ← Keep only top 3 combinations (Output of linear programming formulation)
        SelectedCombination ← Top 1 combination (Output of linear programming formulation)
        Remove/Modify/Add flows on the switches
    end if
end while
```

*D. Scalability Analysis*

The load on our algorithm can be scaled up by increasing the size of the network and the number of flow requests. We will store only top three combinations in *CombinationList* regardless of the network size and the load growth. In this way the complexity of the algorithm in terms of number of combinations will be independent from the number of flow requests. However, the size of each combination will grow as it includes the path set for all the flows.

The algorithm has two phases. First, combinations will be added based on the number of possible paths. There will be $(3 * $ number of possible paths for the new flow$)$ new combinations, where later the top three will be selected and stored. Second, each new combination needs to be updated in terms of its performance metrics. The execution time for the first phase of the algorithm is $O(n)$, where $n$ denotes the number of possible paths for each step. Since $n$ is usually a small number (especially in the FatTree networks) in data center networks and does not grow rapidly along with the network size, it is reasonable to run the algorithm upon each flow request. The second phase of the algorithm updates the performance metrics of a subset of the active switches that will carry the incoming load and it executes in $O(1)$. For each new possible path, active switches can be updated/modified in constant time, which makes our algorithm very scalable with the number of flow requests and the size of the network.

IV. DESIGN OF EVALUATION DECISION FRAMEWORK

We design an evaluation decision framework, whose task is to deploy the different variations of the scheduling algorithm. The decision framework collects the statistical information from the network and schedules new flows or reschedules the existing flows based on online decisions by the LP algorithm. Fig. 1 presents the architecture of the decision framework consisting of three main modules: 1) Scheduler, 2) Controller, and 3) Monitor.

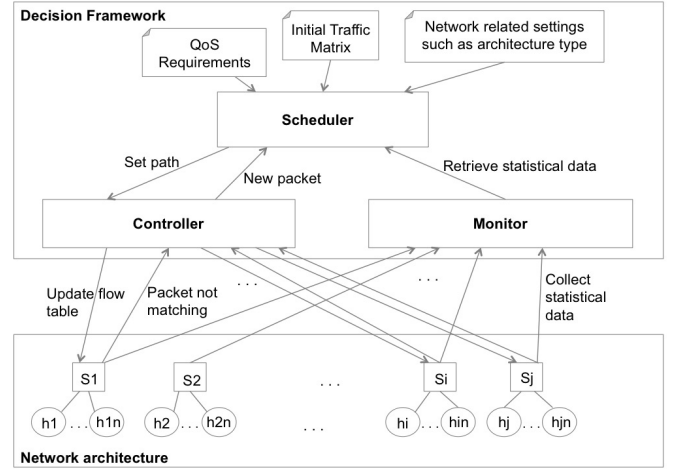

Fig. 1: Our evaluation decision framework consisting of the scheduler, the controller and the monitor components

*A. Scheduler*

The scheduler is at the center of our framework. All the information from other modules and external data sources are inputs to this module. The scheduler deploys the flow scheduling algorithm and it relies on two types of inputs: offline, and online. Monitoring data on power consumption and bandwidth of switches are online inputs. Offline input is given to the module at the initial time and will not change during the runtime. The offline inputs by the scheduler are:

- *QoS requirements*: An application will provide the minimum quality requirements to perform as expected. For example an application might define boundaries for bandwidth and latencies, which will be added as constraints to our linear programming approaches.
- *Initial traffic matrix*: It is possible to provide the scheduler with the traffic setup at the initial time. In this case all flows are fully scheduled at the beginning. We did not use this feature in our experiments.
- *Network architecture*: The scheduler is told at the beginning which network topology is deployed in the data center. Our scheduler can be used to support a number of known network architectures. In our experiments, we use the FatTree topology.

*B. Controller*

The controller component receives the requests for setting new flows from the underline devices. Any time a new packet arrives to a switch, for which an action is not known (the "Packet not matching" function), the packet is sent to the controller component and from there to the scheduler component. Decisions regarding the flow paths are made by the scheduler module, which are communicated later through the "Set path" function to the controller component. The controller, which performs as a middleware between the switches and the scheduler, applies modifications to the flow tables of the switches (the "Update flow table" function).

## C. Monitor

This module collects periodically statistical information from the network (the "Collect statistical data" function) and provides the scheduler with the current status of the flows and utilization of the network devices (the "Retrieve statistical data" function). Requesting stats data from the switches on a timely basis adds overhead in terms of performance degradation and delay increase. It is important to select the right sample rate for switches to report on their flow stats in such a way that no considerable information will be missing and overhead is still tolerable. Reports provided by the monitor are further used by the scheduler component. Other than performance-related data, energy-related information can also be collected by the monitor. The information provided by this module are the input for power consumption models. If realtime or actual monitors are not present, it is possible to calculate the power consumption of the switches based on formalized assumptions. This is what we did in our experiments.

Realtime and online inputs are produced by the network devices and are passed to the scheduler through the controller and the monitor components during runtime.

## V. SIMULATION SCENARIOS

We implemented a number of experiments in a simulation environment using our decision framework. To simulate the "Network Architecture" in Fig. 1 we use Mininet[1], which provides us with a network testbed to develop the software-defined networking system. For the controller component we use the open-source POX control software[2], which updates the flow tables of the simulated Open vSwitches. We implement a combination of the Gurobi Python optimizer[3] and POX as our optimization solver in the scheduler component..

As shown before [3], FatTree is the most widely deployed network architecture in data centers and it is designed to fade out the problem of "single point of failure". FatTree places the switches in a three-level hierarchy, namely core, aggregation and edge switches. Edge switches are also known as top of rack switches (ToR). Aggregation and edge switches form together cells called PODs. We adopted the FatTree topology in our simulations and we use PODs to simulate distant traffic scenarios. We define network architectures of variable sizes namely, 20, 45 and 80 switches all connected with links of 1Gbps, summarized in Table I.

TABLE I: The three simulation configurations of the FatTree network architecture used in our simulation

| Switch ports | Switches | Hosts | PODs | Maximum Flows |
|---|---|---|---|---|
| 4 | 20 | 16 | 4 | 8 |
| 6 | 45 | 54 | 6 | 27 |
| 8 | 80 | 128 | 8 | 64 |

### A. Performance Metrics

Performance metrics help us evaluate through our experiments the energy savings that can be achieved using different variations of the scheduling algorithm and the degradation of other quality requirements. We focus on two performance metrics:

- *Power consumption*: Since we deploy our experiments in a simulation environment, we estimate the power consumption of the data center network. We adopted this model in our simulations:

$$Power_{switch} = Power_{chassis} + \\ num_{linecards} * Power_{linecard} + \\ \sum_{i=0}^{configs}(Power_{configs i} * \sum_{j=0}^{numports} utilizationFactor_j) \quad (10)$$

This is based on the model proposed by Mahadevan *et al.*[19]. In Eq. 10, $Power_{linecards}$ represents the power consumption of the linecard and $num_{linecards}$ is the number of plugged-in cards. $Power_{configs i}$ is the power consumption of a port with the specified link rate $i$ and $utilizationFactor_j$ is the utilization of port $j$ of the switch.
In our experiment, we assume that each flow sends data with the highest possible bandwidth. Therefore, it is always possible to calculate the utilization rates of the switch ports.

- *Time to complete*: We define *time to complete (TTC)* as the time it takes for a host to send predefined number of bytes to another host in the network. TTC is a representative QoS requirement because it is aligned with the response time and the performance.

### B. Traffic Patterns

Depending on the use of the data center, the incoming traffic can follow certain patterns, as identified by [20], which distinguish enterprise and university data centers from cloud data centers based on their running applications. In our current work, we generate the traffic in the data center network based on *One-to-One*, modeled by having all the hosts in pairs [17], [21].

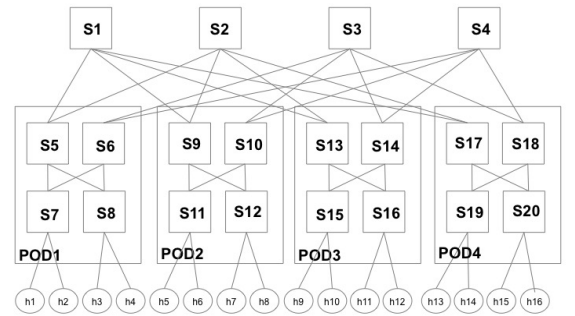

Fig. 2: A FatTree network architecture with 20 switches (S1-S20) and 16 hosts (h1-h16)

As described in [22], the network traffic is categorized based on length of the associated paths (number of switches): 1) Far, 2) Middle, and 3) Near. The latter two involve only one POD respectively with 3 switches and 1 switch. We implement the *Far* traffic study case, where nodes from different PODs transfer data to each other. Since the Far traffic involves more number of switches and less switches can be turned to the sleeping mode, this provides more interesting scenario to validate our algorithms. However, we can utilize the network links fully and investigate more options for the traffic consolidation. For example in Fig. 2, $h1$ connects to a randomly selected host from POD2, POD3 or POD4 and the procedure is continued until all the hosts have been paired.

---
[1] http://mininet.org
[2] http://www.noxrepo.org/pox/about-pox/
[3] http://www.gurobi.com/

## VI. RESULTS

We ran our simulations in the three *FatTree* network architectures identified in Table I, namely with 20, 45 and 80 switches.

In order to evaluate our algorithm variations in terms of power saving and TTC degradation, we first defined the minimum and maximum values of these performance metrics and identified the baseline algorithm we will compare against. A first possible candidate is the shortest-path (SP) algorithm. The drawback of SP is that it always selects the first possible shortest-path for a new flow request and it might schedule many flows on the same link degrading the overall performance. A second possible candidate is a non-energy efficient version of our LP-v3, which we call Smart SP. Smart SP does not put idle devices into the sleeping mode, hence it is not energy efficient, but it maximizes the bandwidth achievements of all the flows and does not select only the first possible path.

We compared SP and Smart SP algorithms in the three network sizes. Table II summarizes the power consumption of data center network of variable sizes when running each of the two algorithms with 100% utilization rate in the one-to-one traffic scenario. As expected, the power consumption of SP and Smart SP is in the same range because they do not put the idle switches into sleeping mode and effectively the power consumption is given by the number of active switches. The little variations we observe between the two algorithms is due to different utilization factors that appear in Eq. 10 given the path chosen will not be always the same.

TABLE II: Total power consumption of SP and Smart SP in the three simulated network topologies (20,45 and 80 switches)

| Baseline candidates | Total power consumption (20 switches) | Total power consumption (45 switches) | Total power consumption (80 switches) |
|---|---|---|---|
| SP | 3032W | 6856W | 12114W |
| Smart SP | 3039W | 6847W | 12198W |

Fig. 3 shows the TTC metric for the two candidate baseline scheduling algorithms in the three simulated network topologies with 100% utilization rate of the one-to-one traffic scenario. In all the 20, the 45 and the 80 switches topologies the TTC of the Smart SP algorithm is lower than the one of the SP algorithm. In fact Smart SP, which makes intelligent decisions regarding bandwidth achievements, effectively ends up providing higher bandwidths to the requesting flows.

Given their equivalent power consumption and the better TTC of Smart SP, we adopt this algorithm rather than SP as baseline algorithm. We expect that the power consumption of our four variations of LP will all improve on Smart SP; at the same time we are interested in quantifying what is the degradation in the TTC in each of the four variations.

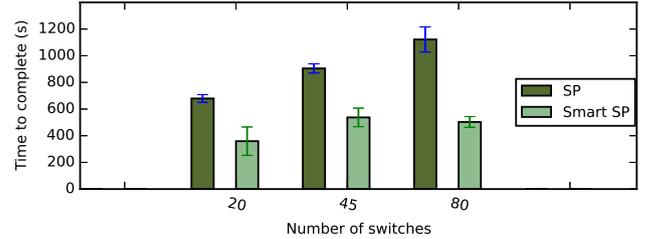

Fig. 3: Time to complete for SP and Smart SP scheduling algorithms in the three simulated network topologies (38GB of data)

It must be noted that power consumption will increase if the load goes up. This increase could be small when due to higher link utilization rates or large when turning on the switches from the sleeping mode. Power consumption values can not be the exclusive metric used to assess the functionality of the scheduling algorithm. Instead, power saving provides a better benchmark among different algorithms: power saving is the amount that the each of scheduling algorithms will save if there is some room for improvement.

The achievable power saving in the sleeping mode will heavily influence the total power saving in the data center. To quantify this, we first examined the power consumption of the four scheduling algorithms under different power savings percentages in sleeping mode. Our switches could save between 20%, 40%, 60% and 80% when in this state, where the 40% and the 60% are the most realistic values. In the rest of this paper we will use 60% as the power saving of the devices in the sleeping mode [23]. Table III shows the maximum power saving of each algorithm compared to the Smart SP in the network of size 45 with 27 flows. In all cases, the maximum power saving is achieved by LP-v4, followed by LP-v1, LP-v2 and LP-v3. There is a nearly linear relation between the maximum achievable power saving and the maximum sleeping mode power saving. 20% improvement of power saving in sleeping mode results in 10% improvement for power sensitive algorithms (LP-v1 and LP-v4) and 5% improvement for performance sensitive algorithms (LP-v2 and LP-v3).

TABLE III: Maximum power savings as function of sleeping mode power savings (45 switches with 27 flows)

| Algorithm variations | Maximum sleeping mode power savings: | | | |
|---|---|---|---|---|
| | 20% | 40% | 60% | 80% |
| LP-v1 | 12% | 23% | 35% | 46% |
| LP-v2 | 6% | 12% | 17% | 23% |
| LP-v3 | 5% | 11% | 16% | 22% |
| LP-v4 | 13% | 24% | 36% | 48% |

To assess the scalability quality requirement of the algorithms we investigated the power consumption of the four algorithm variations when increasing the flow numbers. Fig. 4 shows the results for a network of size 20, 45 and 80 respectively. In the network of size 20 we measure the power consumed in presence of 1, 3, 5 and 8 flows; we used 2, 5, 10, 15, 20 and 27 flows in the network of size 45 and 5, 20, 35, 50 and 64 flows in the network of size 80. In all cases LP-v4 and Smart SP present the lowest and highest values, and as such they identify the lower and highest bound for the power consumption.

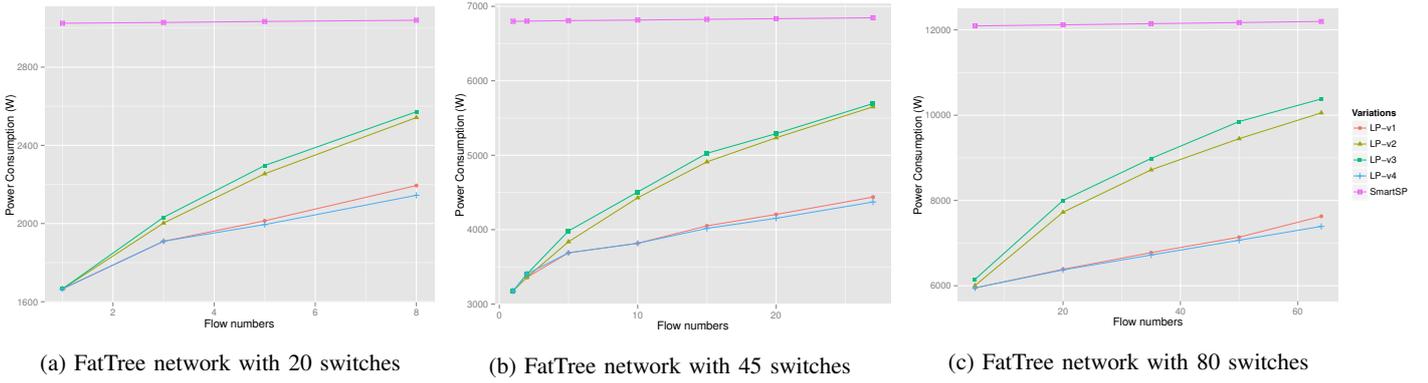

(a) FatTree network with 20 switches   (b) FatTree network with 45 switches   (c) FatTree network with 80 switches

Fig. 4: Power consumption of the data center network for different algorithm variations with variable number of flows based on the one-to-one traffic scenario for a fixed network size

The three subfigures show that LP-v1 selects paths such that the total power consumption remains close to the optimum energy efficient variation (LP-v4). Gradually, when the number of flows increase, LP-v1 and LP-v4 diverge as LP-v1 needs to take into account the throughput requirements too. LP-v2 and LP-v3 show almost identical patterns when increasing the number of flows. Similar to square-root functions, they change rapidly in terms of power consumption for the small number of flows particularly in case of 45 and 80 switches. Their power consumption shows less acceleration for larger number of flows because there are no more switches to turn on from the sleeping mode.

When the number of flows is small all the algorithm variations show the same power consumption; in this case it will not be possible to provide power savings from shaping the network traffic and increasing the number of idle devices. However, small number of flows will involve small number of switches and power savings can be achieved by putting the existing idle devices into the sleeping mode.

Fig. 5 shows the TTC measured from the applications running in the hosts as function of the four possible scheduling algorithms when running with maximum number of flows. We configure applications to send 38GB of data to the receiver. Algorithm variations that perform better in power consumption exhibit higher TTC measurements, e.g. LP-v4 that focuses purely on the energy efficiency will produce a much larger *time_to_complete*, 71% for the network of 20 switches and 502% for the network of 80 switches. LP-v1 also shows a considerable increase in TTC when the network size grows. Differently, LP-v2 and LP-v3 appear to be more performance-focused and stable for different sizes of the network.

If performance is the decisive factor LPv2 and LPv3 will be the likely choice. In this case it is interesting to quantify the degradation in power savings. Fig. 6 shows the degradation in power saving compared to the baseline values (LP-v4 and Smart SP) when using different variations of the scheduling algorithm, compared to the optimal power saving achieved with LPv4. Degradation percentage is calculated from the following, given that $x$ is the power consumption of the algorithm variation:

$$\text{Degradation Percentage}_x = \left(\frac{PC_x - PC_{\text{LP-v4}}}{PC_{\text{Smart SP}} - PC_{\text{LP-v4}}}\right) * 100 \quad (11)$$

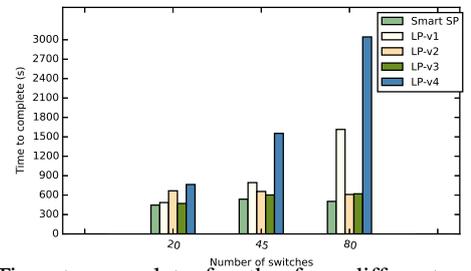

Fig. 5: Time to complete for the four different scheduling algorithm in the three simulated network topologies (38GB of data)

In all three topologies we observed that LP-v1 is the most power efficient variation that shows around 95% improvement, which means only 5% degradation from the optimum power saving. LP-v2 with achieving 50% of maximum power saving outperforms LP-v3 with achieving 45% of maximum power saving, which provides more bandwidth for the running applications. It is interesting to see that all the variations remain with the same range of power saving for different network sizes.

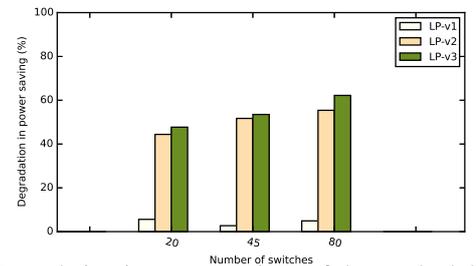

Fig. 6: Degradation in power savings of three scheduling algorithms namely, LP-v1, LP-v2 and LP-v3 in the three simulated network topologies. LP-v4 and Smart SP are considered as the baselines for calculation of power saving.

## VII. DISCUSSION

Applications running in data centers have different quality requirements. Our results can be used as inputs for them to decide on which algorithm to implement for flow scheduling in software-defined networks. As shown in section VI, our four algorithm variations have different behavior as function of the size of the network, the number of flows to be scheduled, and the expected power saving of the switches when in sleeping mode. LP-v2 and LP-v3 provide higher bandwidth for applications and smaller time to complete, while still saving power compared to the shortest-path algorithm. This makes them a great choice for delay-sensitive applications. LP-v1 and LP-v4 focus on the energy efficiency quality requirement rather than the time to complete of applications, as such they can be adopted for delay-insensitive applications. In a data center we will often see a combination of delay-sensitive and delay-insensitive applications. In this case, the scheduler component in our framework can be easily extended to deploy multiple variations of linear programming algorithm concurrently.

Our findings show a nearly linear relation between the total power savings in data center network and the power saving of switches in sleeping mode. It is important to assess the switches power saving in their sleeping mode beforehand, which helps the scheduling algorithms adjust their decisions. Switches with lower sleeping mode power savings are prioritized to carry the incoming load either by performance sensitive algorithms (LP-v2 and LP-v3) or power sensitive algorithms (LP-v1 and LP-v4).

## VIII. CONCLUSION

In this paper we presented different variations of linear programming scheduling algorithms that can optimize the power savings in software-defined networks. Each variation has a different objective in terms of power optimization and performance-related quality requirements. Each objective function implements a combination of following approaches: "put the idle devices into sleeping mode", "increase the number of idle devices", "prioritize the existing active switches over the sleeping ones". The programmability of software-defined networks empowers applications to choose the best flow set and apply direct modifications on the flow tables of the switches. Our algorithm variations are designed for data center networks, in which realtime and scalable traffic scheduling is crucial. We provide scalability by keeping the top 3 flow set candidates, which saves on computation time as new flows come in. We evaluated our algorithm variations in a simulated FatTree network architecture using a one-to-one traffic scenario. We also quantified the relation between power saving of each algorithm and sleeping mode power saving. Our results show that two of the variations (LP-v2 and LP-v3) remain stable in terms of power saving and the time to complete metrics, as the network size grows. Two other (LP-v1 and LP-v4) provide the highest power savings in the network and they are suitable for delay-insensitive applications.

There are a few open questions that we plan to cover in future work. Firstly, many applications will exhibit a one-to-many traffic pattern. In this case, trade-off between power savings and time to complete of the four algorithms might differ from the one-to-one scenario we currently examined. We can expect in general that less switches can be put in the sleeping mode because of the more shared links. Consequently, the variations in power savings between algorithms will be lower. Furthermore, our current implementation of the scheduler treats all flows equally in terms of their expected throughput requirement. In reality, short-lived flows will concurrently run with long-lived flows, as well as high-throughput flows with low-throughput flows. Therefore, it would be interesting to distinguish between them and this can be done by online or offline learning of software traffic patterns.

ACKNOWLEDGMENT

This work has been sponsored by the RAAK/MKB-project "Greening the Cloud" and by the Dutch national program COMMIT.